\documentclass[aps,reprint,prl,showpacs,superscriptaddress]{revtex4-2}
\usepackage{amsmath}
\usepackage{amssymb}
\usepackage{graphicx}
\usepackage{color}
\usepackage{bm}
\usepackage{siunitx}
\usepackage{babel}
\usepackage{hyperref}

\begin{document}

 \title{Transients in shear thickening suspensions: \\when hydrodynamics matters}

\author{Shivakumar Athani}
\affiliation{Univ. Grenoble-Alpes, CNRS, LIPhy, 38000 Grenoble, France}
\author{Bloen Metzger}
\author{Yoël Forterre}
\affiliation{Aix  Marseille  Univ.,  CNRS,  IUSTI,  13453  Marseille,  France}
\author{Romain Mari}
\affiliation{Univ. Grenoble-Alpes, CNRS, LIPhy, 38000 Grenoble, France}

\date{\today}

\begin{abstract}
Using particle-based numerical simulations performed under pressure-imposed conditions, we investigate the transient dilation dynamics of a shear thickening suspension brought to shear jamming.  We show that the stress levels, instead of diverging as predicted by steady state flow rules, remain finite and are entirely determined by the coupling between the particle network dilation and the resulting Darcy backflow. System-spanning stress gradients along the dilation direction lead to cross-system stress differences scaling quadratically with the system size. Measured stress levels are quantitatively captured by a continuum model based on a Reynolds-like dilatancy law and the Wyart-Cates constitutive model. Beyond globally jammed suspensions, our results enable the modeling of inhomogeneous flows where shear jamming is local, e.g. under impact, which eludes usual shear thickening rheological laws.
\end{abstract}


\maketitle

\begin{figure}
\centering
\includegraphics[width=\columnwidth]{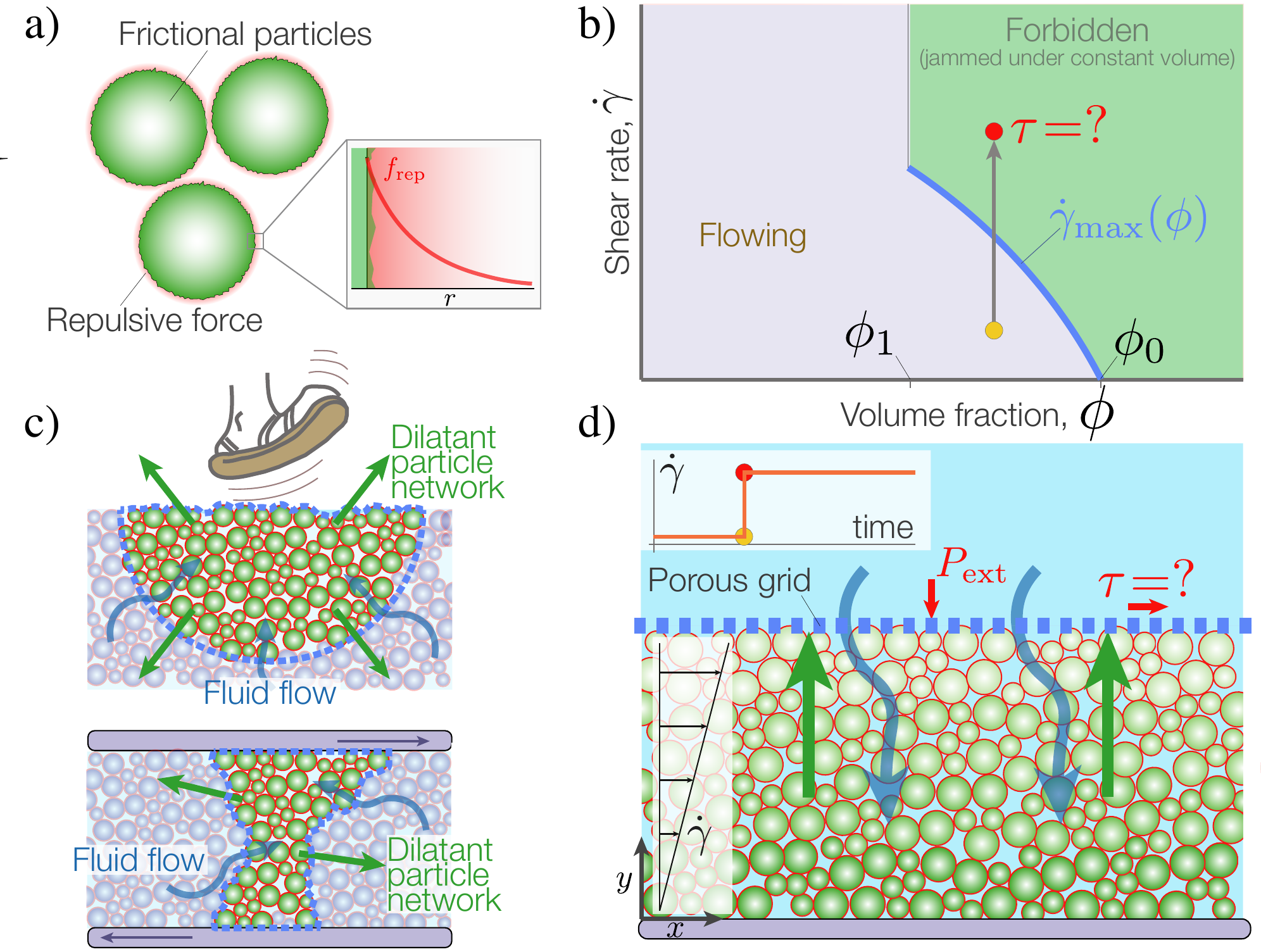}
\caption{ Origin of stress during shear jamming of a shear-thickening suspension. a) Sketch of the frictional and repulsive particles. b) Phase diagram [$\dot\gamma$--$\phi$] shear rate/volume fraction of a shear thickening suspension under constant volume conditions. There is no steady state above the $\dot\gamma_\mathrm{max}(\phi)$ line, which delineates the `forbidden' region (in green). c) Examples of transient shear jamming during impacts and rheometric measurments. d) Model configuration investigated: a shear thickening suspension under imposed external pressure $P_{\rm{ext}}$ is driven in the `forbidden' region by imposing a step increase in shear rate.}
\label{fig1}
\end{figure}

The abrupt shear thickening of dense non-Brownian suspensions is now understood as a frictional transition ~\cite{Fernandez_2013,Seto_2013a,Heussinger_2013,Mari_2014,lin_hydrodynamic_2015,comtet_pairwise_2017,clavaud_revealing_2017,morris_lubricated--frictional_2018,etcheverryCapillaryStressControlledRheometer2023}. This transition occurs when suspended particles interact through a short range repulsive force that prevents solid frictional contacts below a typical applied stress $P_\mathrm{rep}=f_{\rm{rep}}/a^2$, where $f_{\rm{rep}}$ is the amplitude of the repulsive force and $a$ the particle diameter [Fig. \ref{fig1}a]. 
Particles then virtually behave as if they were frictionless, yielding a jamming volume fraction $\phi_0$  significantly larger than the critical packing fraction $\phi_1$ of an assembly of frictional particles. 
Activating frictional contacts by increasing the applied stress above $P_\mathrm{rep}$, then leads to an increase in the suspension viscosity~\cite{Seto_2013a}. 

The most dramatic and interesting case is when the suspension volume fraction $\phi$ is prepared such that $\phi_1<\phi<\phi_0$. The frictional transition then leads to shear-jamming: the frictionless suspension can flow under small stresses [Fig. \ref{fig1}b, orange circle], but jams above a critical stress, which implies that there is a maximum achievable shear rate $\dot\gamma_\mathrm{max}(\phi)$ [Fig. \ref{fig1}b, red circle]~\cite{Wyart_2014,mari_nonmonotonic_2015}.  This situation is encountered during impact~\cite{leeBallisticImpactCharacteristics2003,egresStabResistanceShear2006,waitukaitisImpactactivatedSolidificationDense2012,brassardViscouslikeForcesControl2021}, but also in rheometric measurements in the thickened state, when intermittency arises from the emergence of transient localized jammed structures, see examples shown in Fig. \ref{fig1}c and ~\cite{ovarlezDensityWavesShearthickening2020,ratheeLocalizedTransientJamming2020,gauthierShearthickeningPresenceAdhesive2023,moghimiStressFlowInhomogeneity2024}. Unfortunately, conventional constant-volume constitutive models based on the frictional transition~\cite{nakanishiFluidDynamicsDilatant2012,Wyart_2014,singhConstitutiveModelSimple2018,guyTestingWyartCates2019,mariForceTransmissionOrder2019} cannot predict the stress levels in this case: imposing a non-vanishing shear rate to a jammed system is simply impossible in their framework.  

However, the above scenario only applies under strict volume-imposed conditions.  In practice, jammed particulate systems can  flow under any imposed shear rate, provided that the particle phase is allowed to dilate~\cite{reynoldsDilatancyMedia1885}, for example by deforming the capillary interface or by dilating into the surrounding unjammed suspension.  In the presence of an interstitial fluid, dilation of the particle network induces a Darcy back flow, whose drag on the particles modifies the inter-particle contact forces and thus the stress level. This pore-pressure feedback mechanism has been extensively studied in the context of soil mechanics or geophysics, and has been shown to control the transient flow dynamics of fluid-saturated granular packing, prepared above or below their critical volume fraction~\citep{iversonAcuteSensitivityLandslide2000,pailhaInitiationUnderwaterGranular2008,kulkarni2010particle,rondon2011granular,topin2012collapse,metzger2012clouds,jerome2016unifying,bougouin2018granular,montella2021two,athaniTransientFlowsMigration2022}. Shear thickening suspensions are expected to be particularly prone to such a feedback. First, they can be easily prepared above their jamming volume fraction by increasing the shear rate above $\dot\gamma_\mathrm{max}(\phi)$, and second, the small particle size and the large expected volume-fraction change during dilation should induce particularly strong Darcy pressure gradients.  These effects have never been investigated in shear thickening suspensions, although previous studies suggest they may play an important role \cite{jerome2016unifying,ratheeStructurePropagatingHighstress2022,bougouin2024frictional,talonPressuredrivenPoiseuilleFlow2024}. The aim of the present study is to fill this gap.

In this Letter, we clarify what controls the stress levels during transients in a dilating shear thickening suspension. In order to avoid the spatial and temporal complexities encountered during impact and/or rheometric measurements, while keeping the key physical ingredients, we use the simple shear configuration shown in Fig.~\ref{fig1}d. Discrete numerical simulations are performed under pressure-imposed conditions and the suspension, initially prepared such that $\phi_1<\phi<\phi_0$ [Fig. \ref{fig1}b, orange circle], is suddenly forced into the `forbidden' phase  by imposing a step in shear rate [Fig. \ref{fig1}b, red circle].  We show that during this transient, the resistance to flow of the suspension is entirely set by the coupling between the dilation dynamics and the Darcy backflow. A simple two-phase model, based on Wyart-Cates (WC) flow rules augmented with a Reynolds-like dilatancy law, quantitatively captures the stress level during dilation. The proposed model overcomes the dead-end that pure constant-volume constitutive models face in situation where shear jamming occurs, and highlights an overlooked role of hydrodynamics in setting stresses in non-homogeneous flows of shear-thickening suspensions. 

We simulate a monolayer of $N=2000$ neutrally buoyant non-Brownian hard spheres immersed in a Newtonian fluid of viscosity $\eta_0$, placed between two rigid walls made of frozen particles separated by a mean distance $H$. We apply periodic boundary conditions along the horizontal ($x$) direction.  The top wall is permeable to the fluid and subject to an externally applied normal stress $P_\mathrm{ext}$, but otherwise free to move vertically along the $y$ direction. 
The bottom one is fixed and not permeable.  
A uniform shear $\dot\gamma$ is applied by prescribing a top wall horizontal velocity $\dot\gamma H$. A background fluid velocity field $\bm{v}^\infty(y) = (\dot\gamma y, 0)$  acts on the particles via a Stokes drag $6\pi \eta_0 a_i \bm{v}^\prime_i$, where $\bm{v}^\prime_i = \bm{v}_i - \bm{v}^\infty(y_i)$ is the non-affine velocity of particle $i$, $\eta_0$ the fluid viscosity and $a_i$ the particle radius. This drag force imposes first, a uniform particle shear rate across the layer and second, it sets the Darcy back flow during dilation when particles move vertically (along the direction transverse to the flow)~\citep{athaniTransientFlowsMigration2022}. Particles are bidisperse, with a size ratio $1.4$ and mean radius $a$, mixed in equal volume. 

To simulate a shear thickening suspension, we use frictional particles of friction coefficient $\mu_\mathrm{p}=0.5$ interacting via a zero-range repulsive force $f_{\rm{rep}}$ using the critical load model~\cite{Mari_2014}. We assume a Stokes flow, and neglect inertial effects, so  the equation of motion consists of force balance,
\begin{equation}
   \forall i : \qquad 0 = \sum_j \bm{f}_{\mathrm{c}, ij} - 6\pi \eta_0 a_i \bm{v}^\prime_i + \sum_j \bm{f}_{\mathrm{lub}, ij},
\end{equation}
and torque balance, which takes similar form~\cite{Mari_2014,SuppMat}.
Here $\bm{f}_{\mathrm{c}, ij}$ is the contact force from $j$ on $i$, following $|\bm{f}^\mathrm{t}_{\mathrm{c}, ij}| \leq \mu_\mathrm{p} \max(|\bm{f}^\mathrm{n}_{\mathrm{c}, ij}|-f_\mathrm{rep}, 0)$ 
with $\bm{f}^\mathrm{t}_{\mathrm{c}, ij}$ and $\bm{f}^\mathrm{n}_{\mathrm{c}, ij}$ the tangential and normal parts of the contact force and $f_\mathrm{rep}\equiv P_\mathrm{rep}a^2$ the critical load which defines the thickening pressure scale~\cite{Mari_2014,SuppMat}.  
In the present configuration, the control parameters are the initial system size $H/a$, the viscous number $J_\mathrm{ext} \equiv \eta_0\dot\gamma/P_\mathrm{ext}$ and the reduced pressure $\hat{P}_\mathrm{ext} \equiv P_\mathrm{ext}/P_\mathrm{rep}$, whose value sets the particles frictional state in steady state~\cite{Wyart_2014,dong_analog_2017,etcheverryCapillaryStressControlledRheometer2023}.
The steady-state rheological flow rules of the suspension [$\mu_{\rm{st}}(J_\mathrm{ext},\hat{P}_\mathrm{ext})$ and $\phi_{\rm{st}}(J_\mathrm{ext},\hat{P}_\mathrm{ext})$], where $\mu=\tau/P_p$ is the suspension friction coefficient with $\tau$ the mean shear stress, and  $\phi(t) \equiv N\pi a^2/[l_x (H+\delta H(t))]$ is the solid fraction with $l_x$ the system horizontal length and $H+\delta H(t)$ the top wall position at time $t$, are provided in~\cite{SuppMat}.

\begin{figure}
\centering
\includegraphics[width=\columnwidth]{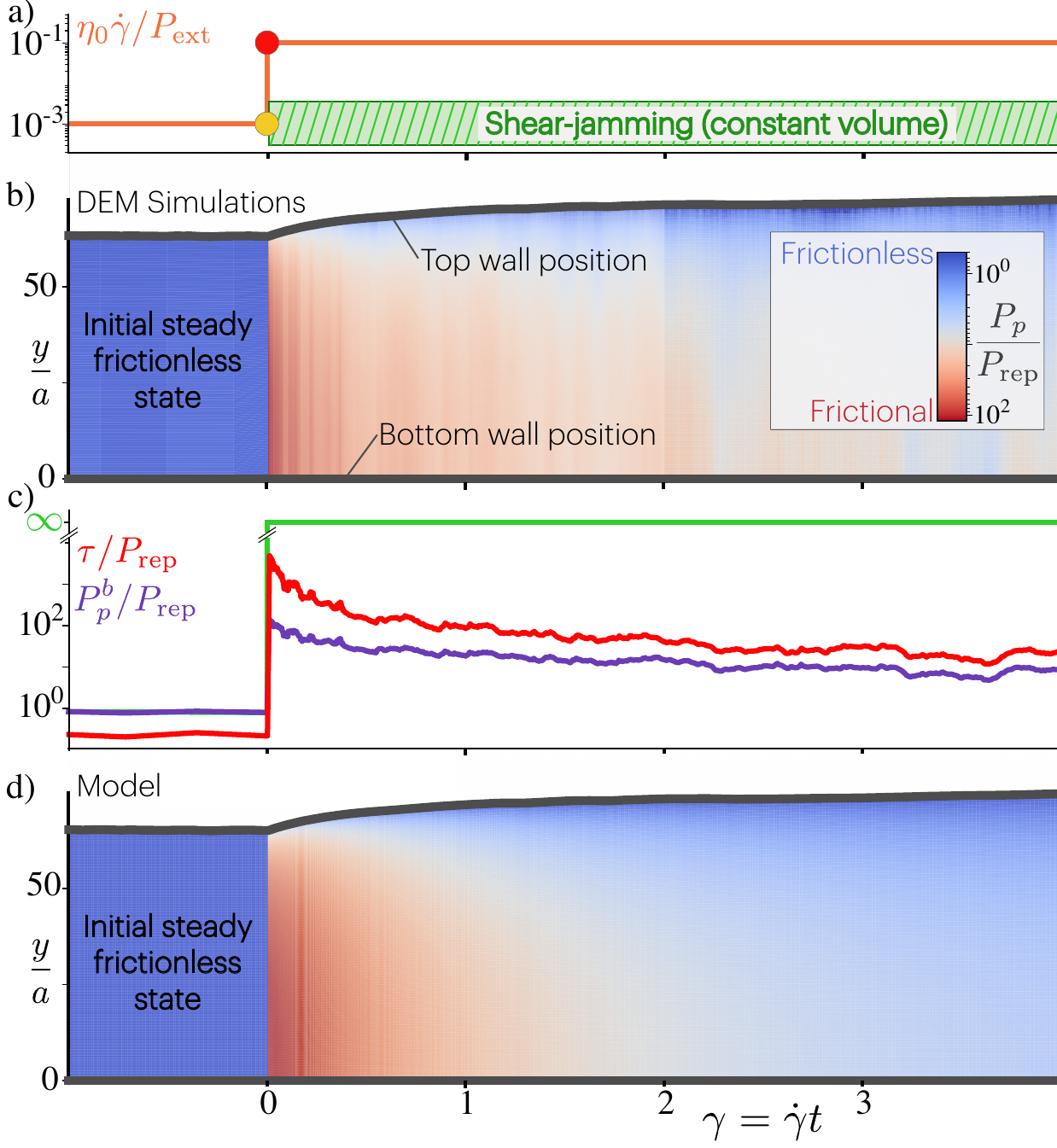}
\caption{ Transient dilation of a shear-jammed suspension yields finite stress.
a) Imposed step increase in shear rate keeping $\hat{P}_\mathrm{ext}=0.75$ constant and choosing $J_{\rm{i}}=10^{-3}$ and $J_{\rm{f}}=10^{-1}$.
Resulting evolution of 
b) the top wall position and of the particle pressure $P_{\rm p}/P_{\rm{rep}}$ (color-coded) within the layer,
c)  average shear stress $\tau$ and particle pressure at the bottom of the layer $P^\mathrm{b}_{\rm p}$, Green solid line: flow rules under constant volume conditions,  
d) model (Eqs. (\ref{eq:SBM}) and (\ref{eq:PP_model})) solved for the same conditions to compare with (b) obtained from DEM simulations.}
\label{fig2}
\end{figure}

The suspension is initially prepared in a frictionless steady-state by setting $\hat{P}_\mathrm{ext} \lesssim 1$ and choosing $J_i$ small enough such that the initial $\phi$ lies between the frictionless and frictional jamming points, $\phi_1 < \phi=\phi_{\rm{st}}(J_i,\hat{P}_\mathrm{ext}\lesssim1) < \phi_0$  (Fig.~\ref{fig1}d, orange circle). At strain $\gamma=0$, while keeping $\hat{P}_\mathrm{ext} \lesssim 1$ constant, we impose a sudden step increase in shear rate, setting $J_\mathrm{f} \equiv \eta_0 \dot\gamma_\mathrm{f}/P_\mathrm{ext} \gg J_\mathrm{i}$, such that if the system was under constant volume, the suspension would instantly show infinite stresses (Fig.~\ref{fig1}d, red circle). 
Instead, the pressure-imposed  configuration used here allows the particle network to dilate, as shown by the evolution of position of the top wall in Fig.~\ref{fig2}b. 
Although the external pressure $P_\mathrm{ext}$ is always kept constant and below $P_{\rm{rep}}$, the vertical dilation of the particle network induces a 
drag on the particles which increases the particles stress $P_{\rm p}$ orders of magnitude above the onset stress to activate frictional contacts. During this transient, the particle pressure $P_{\rm p}$  within the layer is no longer set by the external pressure $P_\mathrm{ext}$ but results from the dilation dynamics. Fig.~\ref{fig2} shows that dilation and the associated Darcy pressure gradient lead to a transient frictional transition occurring in most of the suspension layer (where $P_{\rm p}/P_{\rm{rep}}\gtrsim 10$ colored red  in Fig.~\ref{fig2}b).  At longer times, the suspension reaches a new steady state determined by the values of $J_\mathrm{f}$ and  $\hat{P}_\mathrm{ext}$, and as the effects of dilation/Darcy backflow eventually vanish,  the suspension returns to frictionless.

\begin{figure*}
\centering
\includegraphics[width=17cm]{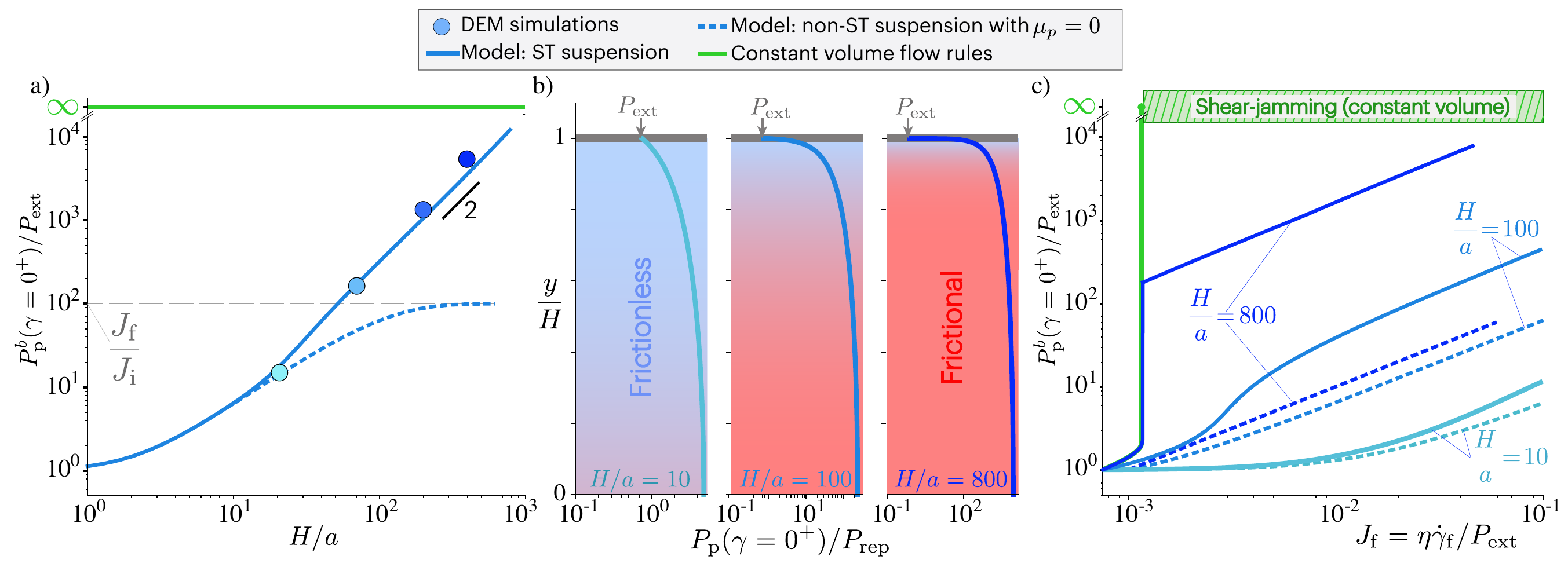}
\caption{ Particle pressure: finite-size and shear-rate effects. Particle pressure  $P_\mathrm{p}^\mathrm{b}(\gamma=0^+)/P_{\rm{rep}}$ at the bottom of the layer and just after the step versus layer thickness $H/a$ ($\hat{P}_\mathrm{ext} = 0.2$, $J_i=10^{-3}$ and $J_f=10^{-1}$ constant). b) Particle pressure profiles  $P_\mathrm{p}(\gamma=0^+)/P_{\rm{rep}}$ (color-coded to highlight the frictionless layer near the top wall) for different system size. c)  Particle pressure $P_\mathrm{p}^\mathrm{b}(\gamma=0^+)/P_{\rm{rep}}$ versus final rescaled shear rate $J_{\rm f}=\eta \dot{\gamma}_{\rm f}/ P_{\rm ext}$ ($\hat{P}_\mathrm{ext} = 0.75$ and $J_\mathrm{i}=10^{-3}$ constant).} 
\label{fig3}
\end{figure*}

These DEM simulations allow us to estimate the particle pressure $P_\mathrm{p}^\mathrm{b}$ at the bottom of the layer (where it is largest) and the average shear stress $\tau$ that result from this step increase in shear rate. While WC flow rules under constant volume would predict an infinite shear stress and particle pressure (green curve in Fig.~\ref{fig2}c), 
the dilation yields finite values with a maximum reached at the onset of dilation, for $\gamma=0^+$ (red and purple curves in Fig.~\ref{fig2}c, respectively). Interestingly, Fig.~\ref{fig3}a shows that, for given values of $J_\mathrm{i}$ and $J_\mathrm{f}$, the particle pressure $P_\mathrm{p}^\mathrm{b}(\gamma=0^+)$ just after the step increases quadratically with the thickness of the layer $H/a$, such that $P_\mathrm{p}^\mathrm{b}/P_{\rm{ext}}\propto (H/a)^2$. We also find that, for the range of parameters investigated, dilation yields particle pressures $P_\mathrm{p}^\mathrm{b}$ reaching values four orders of magnitude larger than the imposed external pressure $P_{\rm{ext}}$. 

These trends can be rationalized as follows. When forced into the forbidden region, we assume that, in order to flow, the suspension dilates uniformly. Mass conservation then prescribes that $\partial_y v^\mathrm{p}_y = - (\partial_t \phi)/\phi = \epsilon$, where $v^\mathrm{p}_y$ is the vertical component of the particle phase velocity due to dilation. The interphase drag force, resulting from the vertical motion of the grains, induces an additional pressure gradient on the particle phase $\partial_y P_\mathrm{p} \propto \eta_0 v^\mathrm{p}_y/a^2$. Altogether, this gives $P_\mathrm{p}^\mathrm{b} - P_\mathrm{ext} \propto \eta_0 \epsilon (H/a)^2$, which for large systems yields $P_\mathrm{p}^\mathrm{b} \propto \eta_0 \epsilon (H/a)^2$, in agreement with Fig.~\ref{fig3}a.

A more detailed two-phase model allows a refined understanding. Mass conservation on the solid phase is
\begin{equation}
\partial_t \phi + \partial_y (v^\mathrm{p}_y \phi) = 0.
\label{eq:massparticle}
\end{equation}
Momentum conservation, in the approximation of the Suspension Balance Model (SBM)~\cite{nottPressuredrivenFlowSuspensions1994,morrisPressuredrivenFlowSuspension1998,morris1999curvilinear,nottSuspensionBalanceModel2011},
balances the pressure gradient $\partial_y P_\mathrm{p}(y)$ with the local interphase drag, which is directly proportional to $v^\mathrm{p}_y$
\begin{equation}
 \partial_y P_\mathrm{p}(y) = - \frac{\eta_0}{a^2} \phi R(\phi) v^\mathrm{p}_y,
 \label{eq:SBM}
\end{equation}
with $R(\phi)$ the hydraulic resistance of the particle network, which is dimensionless in 3D. 
We use $R(\phi)=6a$ to adapt to our 2D simulations with a simple Stokes drag, with $P_\mathrm{p}$ now a force per unit length. The model is closed with a Reynolds-like dilation law~\cite{pailha_two-phase_2009}
\begin{equation}
    \partial_t \phi + v^\mathrm{p}_y \partial_y \phi = - \frac{\dot\gamma}{\gamma_0}\left[\phi - \phi_\mathrm{st}(J, \hat{P}_\mathrm{p})\right],
    \label{eq:PP_model}
\end{equation}
where $\phi$ relaxes towards its steady-state value $\phi_\mathrm{st}(J, \hat{P}_\mathrm{p})$ (with $\hat{P}_\mathrm{p} = P_\mathrm{p}/P_\mathrm{rep}$) on a strain scale $\gamma_0$~\cite{rouxTextureDependentRigidPlasticBehavior1998}.
The steady rheology is given by WC model, where for this shear-thickening suspension, the steady-state solid fraction $\phi_\mathrm{st}(J, \hat{P}_\mathrm{p})$ depends both on $J$ and $\hat{P}_\mathrm{p}$~\cite{Wyart_2014,SuppMat}. Finally, boundary conditions are simply $v^\mathrm{p}_y = 0$ at the bottom wall and $P_\mathrm{p} = P_\mathrm{ext}$ at the top wall.

The evolution of the predicted particle pressure within the layer is shown in Fig.~\ref{fig2}d, which highlights the transient frictional transition induced by dilatancy, in good agreement with the DEM results obtained for the same conditions and shown in  Fig.~\ref{fig2}b.  Furthermore,  Fig.~\ref{fig3}a (blue solid line) shows that the model also captures the $(H/a)^2$ dependence of $P_\mathrm{p}^\mathrm{b}(\gamma= 0^+)$ for large system size. By contrast, for a frictionless (non shear-thickening) suspension ($\mu_\mathrm{p}=0$, blue dashed line), the solid fraction always remains below its jamming value $\phi_0$. 
For large systems, the particle stress just after the step is then simply set by the steady state flow rules, $P_{\rm{p}}^{\rm b}(\gamma=0^+)/P_{\rm{ext}}=J_{\rm{f}}/J_{\rm{i}}$, and is independent of the dilation~\cite{athaniTransientsPressureimposedShearing2021}.  

At this point, we highlight and discuss finite-size effects.
Right after the step, the particle pressure profile $P_{\rm{p}}(\gamma=0^+)$ instantaneously establishes across the layer, see Fig.~\ref{fig3}.b. 
This is because $\partial_t \phi$ appears both in Eqs.~(\ref{eq:massparticle}) and (\ref{eq:PP_model}), so that it can be eliminated to reveal that at any instant the pressure obeys a non-linear second order ODE~\cite{athaniTransientFlowsMigration2022}.
Since the pressure profile must match the boundary condition $P_\mathrm{p}(y=H) =P_\mathrm{ext}$, there is always a region below the bottom wall which remains frictionless and unjammed (where $P_{\rm p}/P_{\rm{rep}}\lesssim 10$). 
Moreover, one also expects that $P^\mathrm{b}_\mathrm{p}(\gamma=0^+) \to P_\mathrm{ext}$ when $H\to 0$. 
Altogether, this explains why, for small systems, the dependence of $P^\mathrm{b}_\mathrm{p}(\gamma=0^+)$ on $H$ is milder than $H^2$.

In Fig.~\ref{fig3}.c, we further explore the influence of the magnitude of the step in shear rate, keeping $\dot{\gamma}_{\rm{i}}$ and $P_{\rm{ext}}$ constant, but varying systematically $\dot{\gamma}_{\rm{f}}$ for different system sizes. As expected from Eq.~(\ref{eq:PP_model}), the particle pressure for large system sizes scales linearly with the imposed shear rate $\propto \dot\gamma_\mathrm{f}$, whenever $\dot{\gamma}_{\rm f}$ exceeds $\dot\gamma_\mathrm{max}(\phi)$, which is
where the constant-volume rheology would predict that $P^\mathrm{b}_\mathrm{p}(\gamma=0^+)$ diverges (green solid line). For smaller systems, the particle pressure is damped, again as a manifestation of the unjammed layer below the top wall.

To conclude, we investigate the stress levels that develop when a shear-thickening suspension is suddenly brought to shear jamming, we perform DEM simulations in a model pressure-imposed configuration where the suspension is allowed to dilate. We show that stresses do not diverge, as expected  for strictly volume-imposed conditions. Instead, stresses remain finite and are determined entirely by the coupling between the Reynolds dilation and the Darcy backflow. 
In the large size limit, this Darcy-Reynolds coupling induces a particle pressure difference across the system, that scales quadratically with the system size. We model and rationalize the above phenomenology with a two-phase migration law taking a Reynolds-dilatancy form~\cite{pailha_two-phase_2009,athaniTransientsPressureimposedShearing2021}, the suspension balance model~\cite{nottPressuredrivenFlowSuspensions1994,morrisPressuredrivenFlowSuspension1998,morris1999curvilinear,nottSuspensionBalanceModel2011}, and WC constitutive flow rules~\cite{Wyart_2014}. 

Our results show that a dilation, even minute, can be the source of very large transient stresses within the shear-jammed region. We anticipate that this mechanism could play an important role in determining the resistance of shear thickening suspensions to impulsive motion, such as during impact, as previously proposed~\cite{jerome2016unifying}. Other mechanisms have been suggested, either based on the inertia of the impactor~\cite{waitukaitisImpactactivatedSolidificationDense2012}, its ability to transmit stress to the boundaries~\cite{vonkannNonmonotonicSettlingSphere2011a,petersQuasi2DDynamicJamming2014,allenSystemspanningDynamicallyJammed2018,mukhopadhyayTestingConstitutiveRelations2018,hanDynamicJammingDense2019}, its elasticity~\cite{pradiptoImpactinducedHardeningDense2021,pradiptoEffectiveViscosityElasticity2023}, or its viscous drag at the boundaries~\cite{brassardViscouslikeForcesControl2021}. Crucially these scenarii only give to the jammed region the role of a rigid stress transmitter, but there is also evidence of dilation-induced stresses  in these systems~\cite{jerome2016unifying}. Our simulations show that the particle pressure resulting from dilatancy reaches up to $\num{e3}P_\mathrm{rep}$ for system sizes $H/a \lesssim 400$. Extrapolating this value to actual flows on the cm scale, with particle sizes $a$ in the \textmu m  range, we expect pressures $>\num{e6}P_\mathrm{rep}$.  For a cornstarch suspension (with $P_\mathrm{rep}\approx 1-10$ Pa ~\cite{fallShearThickeningCornstarch2008,brownDynamicJammingPoint2009,hermesUnsteadyFlowParticle2016}), this yields about $10-100$ bar, which is  sufficient to transiently support the weight of a person. Interestingly, these dilation-induced stress values are also greater than the pressure required for the grains to pop out of the suspension-air interface, which is of the order of $10\Gamma/a$, where $\Gamma$ is the water-air surface tension~\cite{hackett1928capillary}.  We therefore also expect dilation to be a major cause of free-surface fractures, as commonly observed for macroscopic flows of shear thickening suspensions~\cite{brownShearThickeningConcentrated2014}. 

More fundamentally, while the fluid plays no role in the frictional transition  mechanism  at the origin of the steady state flow rules of shear thickening suspensions, we show here that the consideration of hydrodynamics is key to understanding their transient dynamics. Our results could therefore help to better understand the reported spatio-temporal fluctuations, which elude a steady-state description. Discontinuous shear thickening leads to flow instabilities,  often associated to the formation of transient localized structures showing large stress gradients~\cite{nakanishiFluidDynamicsDilatant2012,nagahiroNegativePressureShear2016,hermesUnsteadyFlowParticle2016,ratheeLocalizedStressFluctuations2017,chackoDynamicVorticityBanding2018,saint-michelUncoveringInstabilitiesSpatiotemporal2018,ratheeLocalizedTransientJamming2020,ovarlezDensityWavesShearthickening2020,darboistexierSurfacewaveInstabilityInertia2020,ratheeStructurePropagatingHighstress2022,gauthierShearthickeningPresenceAdhesive2023,moghimiStressFlowInhomogeneity2024,angermanNumericalSimulationsSpatiotemporal2024} and often appearing jammed or rigid~\cite{ratheeLocalizedTransientJamming2020,ovarlezDensityWavesShearthickening2020,gauthierShearthickeningPresenceAdhesive2023}.
Discontinuous shear thickening also correlates with surprisingly large volume fraction fluctuations~\cite{benderReversibleShearThickening1996,chengImagingMicroscopicStructure2011}. 
Steady-state rheological models struggle to account for these fluctuations because of their extreme volume fraction sensitivity: a tiny local increase in volume fraction is quickly smeared out by an outward particle migration from the large increase in local pressure. We show with Eq.~(\ref{eq:PP_model}) that this stabilizing mechanism is not instantaneous, but instead takes a finite strain to act, thus weakening its effect. As it is well known that fluctuations and their dynamics have a key role close to critical points, explicit accounting of transient dilation effects on the dynamics of volume fraction fluctuations near jamming appears as a priority for future developments.

{\em Acknowledgements.} 
This work was funded by ANR ScienceFriction (ANR-18-CE30-0024).
\nocite{Cundall_1979,Luding_2008,jeffreyCalculationLowReynolds1992}

\bibliographystyle{apsrev4-2}
\bibliography{bib}

\end{document}



\title{Supplemental Material\\ \vspace{0.3 cm} Transients in shear thickening suspensions: \\when hydrodynamics matters}

\author{Shivakumar Athani}
\affiliation{Univ. Grenoble-Alpes, CNRS, LIPhy, 38000 Grenoble, France}
\author{Bloen Metzger}
\author{Yoël Forterre}
\affiliation{Aix  Marseille  Univ.,  CNRS,  IUSTI,  13453  Marseille,  France}
\author{Romain Mari}
\affiliation{Univ. Grenoble-Alpes, CNRS, LIPhy, 38000 Grenoble, France}

\date{\today}

\maketitle



\subsection{SM1.~Supplementary movie}
		
	\noindent\textbf{Movie 1:} DEM simulation of a shear thickening suspension forced into shear jamming, by imposing a step in shear rate under pressure imposed conditions. Control parameters: $H/a=400$, $J_{\rm i}=10^{-3}$, $J_{\rm f}=10^{-1}$, $\hat{P}_\mathrm{ext} \equiv P_\mathrm{ext}/P_\mathrm{rep}$=0.75. The frictional state of the particles is color-coded.\vspace{0.2 cm}

\begin{figure}
\includegraphics[width =16 cm]{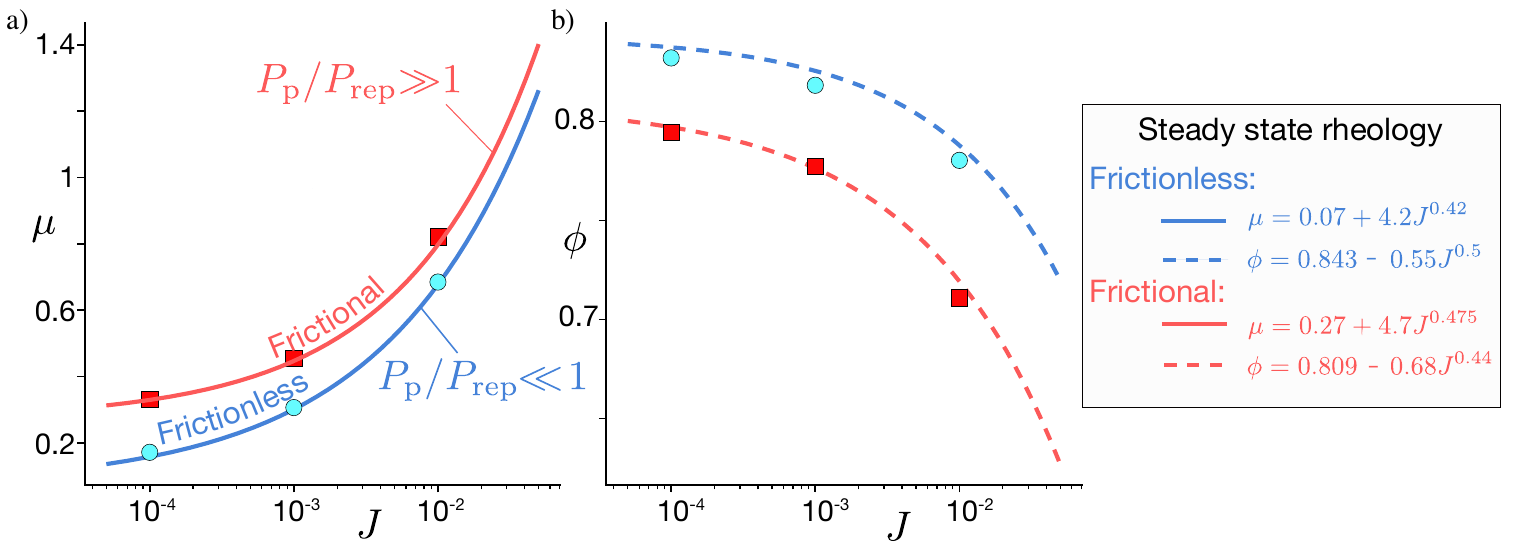}
\caption{{\color{blue} Steady state rheology of the shear thickening suspension.} a) Suspension friction coefficient $\mu$ and b) volume fraction $\phi$ versus $J=\eta \dot{\gamma}/P_{\rm{ext}}$ obtained by setting $P_{\rm p}/P_{\rm{rep}}>>1$ (frictional branch) or $P_{\rm p}/P_{\rm{rep}}<1$ (frictionless branch). Symbols: DEM simulations. Solid lines: best fits used to set Wyart \& Cates model.}
\label{fig:SSrheol}
\end{figure}

\subsection{SM2.~Steady state rheological flow rules}

The steady-state rheology of a shear thickening suspension under imposed particle pressure only depends on the viscous number $J \equiv \eta_0\dot\gamma/P_\mathrm{ext}$ and the reduced pressure $\hat{P}_\mathrm{p} \equiv P_\mathrm{p}/P_\mathrm{rep}$.  Figure \ref{fig:SSrheol} shows the steady state rheology of the suspension obtained from the DEM simulations performed either under low or large imposed external pressure (frictionless or frictional branches, respectively). The best fit of this data is used to set Wyart \& Cates flow rules used in the Darcy-Reynolds model.  

In the limit of low $\hat{P}_\mathrm{p} \ll 1$ (frictionless state), we obtain $\phi = \phi_0 - k_0 J^{\beta_0}$, with $\phi_0 = 0.843$, $k_0 = 0.55$ and $\beta_0 = 0.5$, while at large $\hat{P}_\mathrm{p} \gg 1$ (frictional state), $\phi = \phi_1 - k_1 J^{\beta_1}$, with $\phi_1 = 0.809$, $k_1 = 0.68$ and $\beta_1 = 0.44$~\cite{athaniTransientsPressureimposedShearing2021}. Wyart-Cates flow rules~\cite{Wyart_2014}  is then written as 
$\phi_\mathrm{st}(J, \hat{P}_\mathrm{p}) = \phi_\mathrm{J}(\hat{P}_\mathrm{p}) - K(\hat{P}_\mathrm{p}) J^{\beta(\hat{P}_\mathrm{p})}$,  where $\phi_\mathrm{J}$, $K$ and $\beta$ are functions interpolating between their frictionless and frictional limits. This interpolation is done using the fraction of frictional contacts $f$ as an order parameter, which is defined as $f(\hat{P}_\mathrm{p}) = \exp(-A/\hat{P}_\mathrm{p})$. This yields $\phi_\mathrm{J}(\hat{P}_\mathrm{p}) = f \phi_1 + (1-f) \phi_0$, $K(\hat{P}_\mathrm{p}) = f K_1 + (1-f) K_0$ and $\beta(\hat{P}_\mathrm{p}) = f \beta_1 + (1-f) \beta_0$~\cite{singhConstitutiveModelSimple2018,mariForceTransmissionOrder2019}. 
Only $A\approx 4.5$ is a shear-thickening-specific parameter, and is adjusted to the steady rheology we measure.

\subsection{SM3.~Discrete Element Method equations of motion}

The granular suspension considered in this study is neutrally buoyant and we work in vanishing Stokes number and Reynolds number regime; consequently, the equation of motion is just the force and torque balances on each particle.
Forces include contact forces, lubrication forces, and for the wall particles external forces.
For a pair of particles $i$ and $j$ with radii $a_i$ and $a_j$ at positions $\bm{r}_i$ and $\bm{r}_j$, we note $h_{ij} = 2(|\bm{r}_j - \bm{r}_i| - a_i - a_j)/(a_i + a_j)$ the normalized gap between the particles, and $\bm{n}_{ij} = (\bm{r}_j - \bm{r}_i)/|\bm{r}_j - \bm{r}_i|$ the unit center-to-center vector.

The contact force on a particle \(i\) with radius \(a_i\) in contact with particle \(j\) (i.e., when $h_{ij} < 0$)
can be decomposed in normal and tangential components
\begin{equation}
  \bm{f}_{\mathrm{c},ij} = \bm{f}_{\mathrm{c},ij}^\mathrm{n} + \bm{f}_{\mathrm{c},ij}^\mathrm{t}.
\end{equation}
Contacts fulfill a modified Coulomb's friction laws 
\(\bigl|\bm{f}_{\mathrm{c},ij}^\mathrm{t} \bigr| \leq \max\left[\mu_\mathrm{p} (|\bm{f}_{\mathrm{c},ij}^\mathrm{n}| - f_\mathrm{rep}), 0 \right]\) 
with  sliding friction coefficient \( \mu_\mathrm{p} \).
The introduction of $f_\mathrm{rep}$ ensures that particles are frictionless where interparticle forces are small, and frictional otherwise, ultimately leading to a shear-thickening rheology.
The force components \(\bm{f}_{\mathrm{c},ij}^\mathrm{n}\) and \(\bm{f}_{\mathrm{c},ij}^\mathrm{t}\) 
are modeled in a Cundall-Strack fashion~\citep{Cundall_1979, Luding_2008}, with normal and tangential couples of spring and dashpot
\begin{equation}
  \label{eq:contact_model}
  \begin{split}
    \bm{f}_{\mathrm{c},ij}^\mathrm{n} & = k_{\mathrm{n}} \bm{\xi}_{\mathrm{c},ij}^\mathrm{n} 
     + \gamma_{\mathrm{n}}  \bm{v}^{\mathrm{n}}_{ij},  \\
    \bm{f}_{\mathrm{c},ij}^\mathrm{t} & = k_{\mathrm{t}} \bm{\xi}_{\mathrm{c},ij}^\mathrm{t}
    +\gamma_{\mathrm{t}}  \bm{v}^{\mathrm{t}}_{ij},
  \end{split}
\end{equation}
where $\bm{\xi}_{\mathrm{c},ij}^\mathrm{n} = h_{ij} \bm{n}_{ij}$ is the normal spring stretch, 
$\bm{\xi}_{\mathrm{c},ij}^\mathrm{t}$ is the tangential spring stretch,
$\bm{v}^{\mathrm{n}}_{ij} = (\bm{I} - \bm{n}_{ij}\bm{n}_{ij})\cdot (\bm{v}_j - \bm{v}_i)$ is the normal velocity difference (with $\bm{I}$ the identity matrix),
 $\bm{v}^{\mathrm{t}}_{ij} = \bm{v}_j - \bm{v}_i - \bm{v}_{\mathrm{n}}^{(i,j)} - (a_i \bm{\omega}_i + a_j \bm{\omega}_j)\times \bm{n}_{ij}$ is the tangential relative surface velocity with $\bm{\omega}_i$ and $\bm{\omega}_j$ the angular velocities of particles $i$ and $j$, $k_{\mathrm{n}}$ and $k_{\mathrm{t}}$ are spring stiffnesses and $\gamma_{\mathrm{n}}$ and $\gamma_{\mathrm{t}}$ are dashpot resistances.
Finally, the contact torque on particle $i$ from the contact with particle $j$ is simply obtained as 
$\bm{t}_{\mathrm{c},ij} = a_i \bm{n}_{ij} \times \bm{f}_{\mathrm{c},ij}$. 

Using the notation $\bm{F}_\mathrm{c} \equiv \left(\bm{f}_{\mathrm{c},1}, \dots, \bm{f}_{\mathrm{c},N}, \bm{t}_{\mathrm{c},1}, \dots, \bm{t}_{\mathrm{c},N}\right)$ for the $3N$ vector of contact forces and torques on each particle, we have
\begin{equation}
\bm{F}_\mathrm{c}
=
- \bm{R}^\mathrm{c}_\mathrm{FV} \cdot \bm{V}
+ 
\bm{F}_\mathrm{c,s}
\end{equation}
with $\bm{V} = \left(\bm{v}_1, \dots, \bm{v}_N, \bm{\omega}_1, \dots, \bm{\omega}_N\right)$ the vector of velocities and angular velocities for each particle, $\bm{R}^\mathrm{c}_\mathrm{FV}$ the resistance matrix associated to dashpots, and $\bm{F}_\mathrm{c,s}$ the part of contact forces coming from springs.
For hydrodynamic forces and torques $\bm{F}_\mathrm{h}$, we have~\citep{jeffreyCalculationLowReynolds1992}
\begin{equation} \label{eqhydro}
\bm{F}_\mathrm{h}
=
- \bm{R}^\mathrm{h}_\mathrm{FV} \cdot \bm{V}^\prime
+ 
\bm{R}_\mathrm{FE}:\bm{E}^\infty
\end{equation}
with $\bm{V}^\prime = \bm{V} - \bm{V}^\infty$ the vector of non-affine velocities and angular velocities for each particle, where $\bm{V}^\infty$ are the background velocities and angular velocities evaluated at the particles centers, 
and $\bm{E}^\infty$ the symmetric part of the imposed velocity gradient $\nabla \bm{V}^\infty$.
The resistance matrices $\bm{R}^\mathrm{H}_\mathrm{FV}$ and $\bm{R}_\mathrm{FE}$ contain Stokes drag and torque and regularized lubrication forces and torques at leading order in particle separation~\cite{Mari_2014}.

The equation of motion is thus
\begin{equation}\label{eq:eq_motion_full}
    \bm{F}_\mathrm{h} + \bm{F}_\mathrm{c} + \bm{F}_\mathrm{ext} = 0\, ,
\end{equation}
with $\bm{F}_\mathrm{hxt}$ the external force/torque applied on each particle. Of course, $\bm{F}_\mathrm{ext}$ takes non-zero values only for the wall particles.

As the simulated system comprises of frozen wall and bulk particles suspended in the fluid, we separate the total resistance matrix as
\begin{equation} \label{eqresmatrix}
\bm{R}^\mathrm{c}_\mathrm{FV} + \bm{R}^\mathrm{h}_\mathrm{FV} \equiv \bm{R}_\mathrm{FV}
=
\begin{pmatrix}
\bm{R}_\mathrm{FV}^\mathrm{bb} & \bm{R}_\mathrm{FV}^\mathrm{bw}\\
\bm{R}_\mathrm{FV}^\mathrm{wb} & \bm{R}_\mathrm{FV}^\mathrm{ww}
\end{pmatrix}
\end{equation}
The matrices $\bm{R}_\mathrm{FV}^\mathrm{bb}$, $\bm{R}_\mathrm{FV}^\mathrm{bw}$, $\bm{R}_\mathrm{FV}^\mathrm{wb}$ and $\bm{R}_\mathrm{FV}^\mathrm{ww}$ indicate the hydrodynamic resistance matrices including bulk-bulk, bulk-wall, wall-bulk and wall-wall interactions respectively.
Similarly, the non-affine velocities/angular velocities have also been separated into a bulk particle part $\bm{V}^\mathrm{b} - \bm{V}^{\infty}$ and a wall particle part $\bm{V}^\mathrm{w} - \bm{V}^{\infty}$,
\begin{equation} \label{equdecom}
\bm{V}^\prime 
=
\begin{pmatrix}
\bm{V}^\mathrm{\prime b}\\
\bm{V}^\mathrm{\prime w}
\end{pmatrix}\, .
\end{equation}
Note that the wall particles do not follow the affine flow $\bm{V}^{\infty}$ for three reasons. 
First, their velocity have a vertical component when the system dilates or contracts. 
Second, even the horizontal component of the velocity of a wall particle, say particle $i$, located at height $y_i$, does not match $\bm{U}^{\infty}(y_i)$ because the wall moves rigidly and is made of particles located at slightly different heights.
Third, they are constrained to have vanishing angular velocity, when the affine flow has a finite angular velocity.

We can then solve the equations of motion, Eq.~\ref{eq:eq_motion_full} for the bulk velocities, to get
\begin{equation}\label{eq:bulk_velocities}
    \bm{V}^\mathrm{\prime b} = {\bm{R}_\mathrm{FV}^{\mathrm{bb}}}^{-1} \cdot \left[\bm{K}^\mathrm{b} - \bm{R}_\mathrm{FV}^\mathrm{bw}\cdot \bm{V}^\mathrm{\prime w}\right]
\end{equation}
with
\begin{equation}
    \begin{pmatrix}
    \bm{K}^\mathrm{b}\\
    \bm{K}^\mathrm{w}
    \end{pmatrix} = \bm{R}_\mathrm{FE}:\bm{E}^\infty + \bm{F}_\mathrm{c,s} - \bm{R}^\mathrm{c}_\mathrm{FV} \cdot \bm{V}^\infty
\end{equation}
We further decompose the wall non-affine velocity in horizontal and vertical components $\bm{V}^\mathrm{\prime w} = \bm{V}^\mathrm{\prime w}_\mathrm{h} + v_y \bm{Y}^\mathrm{w}$, where $\bm{Y}^\mathrm{w}$ is the vector corresponding to a unit non-affine vertical velocity  for particles belonging to the upper wall, and vanishing non-affine velocity for particles of the lower wall.
Thus the scalar $v_y = \partial_t \delta H$ is the upper wall vertical speed.

Injecting Eq.~\ref{eq:bulk_velocities} in the wall part of Eq.~\ref{eq:eq_motion_full}, and using Eq.~\ref{equdecom}, we get
\begin{equation}
   v_y \bm{B}\cdot \bm{Y}^\mathrm{w} = \bm{K}^\mathrm{w} - \bm{R}_\mathrm{FV}^\mathrm{wb}\cdot {\bm{R}_\mathrm{FV}^{\mathrm{bb}}}^{-1}\cdot \bm{K}^\mathrm{b} - \bm{B}\cdot \bm{V}^\mathrm{\prime w}_\mathrm{h} + \bm{F}_\mathrm{ext}\, .
\end{equation}
Taking the dot product with $\bm{Y}^\mathrm{w}$, and using $\bm{Y}^\mathrm{w}\cdot \bm{F}_\mathrm{ext} = P_\mathrm{ext} l_x$ we get the vertical wall velocity
\begin{equation}
     v_y = \frac{\bm{Y}^\mathrm{w} \cdot\left[ \bm{K}^\mathrm{w} - \bm{R}_\mathrm{FV}^\mathrm{wb}\cdot {\bm{R}_\mathrm{FV}^{\mathrm{bb}}}^{-1}\cdot \bm{K}^\mathrm{b} - \bm{B}\cdot \bm{V}^\mathrm{\prime w}_\mathrm{h}\right] + P_\mathrm{ext} l_x}{\bm{Y}^\mathrm{w}\cdot \bm{B}\cdot \bm{Y}^\mathrm{w}}\, .
\end{equation}
Imposing that the vertical component of the upper wall velocity is $v_y$ ensures that the total force on the upper wall vanishes, and therefore that the applied pressure on the top wall is indeed $P_\mathrm{ext}$.

\bibliographystyle{apsrev4-1}
\bibliography{../bib.bib}